\documentclass[reprint,prb,twocolumn,longbibliography,noeprint,superscriptaddress,floatfix]{revtex4-2}
\usepackage{float}
\usepackage{physics}
\usepackage{graphicx}
\usepackage{xcolor}
\usepackage{amsmath,amsfonts,amssymb}
\usepackage{textcomp}
\usepackage{bbm}
\usepackage{dsfont}
\usepackage{microtype}
\usepackage{mathtools}
\usepackage{multirow}

\usepackage{array}
\newcolumntype{L}[1]{>{\raggedright\let\newline\\\arraybackslash\hspace{0pt}}m{#1}}
\newcolumntype{C}[1]{>{\centering\let\newline\\\arraybackslash\hspace{0pt}}m{#1}}
\newcolumntype{R}[1]{>{\raggedleft\let\newline\\\arraybackslash\hspace{0pt}}m{#1}}

\usepackage[pdftex,bookmarks=true,bookmarksopen,bookmarksnumbered,
                colorlinks,
                linkcolor=blue,
                citecolor=blue,
                colorlinks = true,
                urlcolor  = blue,
                anchorcolor = blue
                ]{hyperref}
\usepackage{cleveref}

\renewcommand{\figurename}{Figure}
\makeatletter
\renewcommand*{\fnum@figure}{{\normalfont \figurename~\thefigure}}
\renewcommand*{\@caption@fignum@sep}{ $~$}

\renewcommand{\tablename}{Table}
\makeatletter
\renewcommand*{\fnum@table}{{\normalfont \tablename~\thetable}}

\usepackage{float}
\usepackage{newfloat}
\usepackage{standalone}
\usepackage{booktabs}
\usepackage[geometry]{ifsym}

\crefname{figure}{Figure}{Figures}
\crefname{table}{Table}{}
\crefname{section}{Sec.}{Secs.}
\crefname{equation}{Eq.}{Eqs.}

\graphicspath{{paper_figures/}}

\newcommand{\al}[1]{\textcolor{black}{#1}}

\usepackage{lineno}
\begin{document}

\title{\al{Quantum Key Distribution Using a Quantum Emitter in Hexagonal Boron Nitride}}

\author{Ali Al-Juboori}
\altaffiliation[]{These authors contributed equally to this work.}
\affiliation{
School of Mathematical and Physical Sciences, Faculty of Science, University of Technology Sydney, Ultimo, New South Wales, 2007, Australia
}
\affiliation{
School of Electrical Engineering and Telecommunications, The University of New South Wales, Sydney, NSW 2052, Australia
}

\author{Helen Zhi Jie Zeng}
\altaffiliation[]{These authors contributed equally to this work.}
\affiliation{
School of Mathematical and Physical Sciences, Faculty of Science, University of Technology Sydney, Ultimo, New South Wales, 2007, Australia
}

\author{Minh Anh Phan Nguyen}
\affiliation{
School of Mathematical and Physical Sciences, Faculty of Science, University of Technology Sydney, Ultimo, New South Wales, 2007, Australia
}

\author{Xiaoyu Ai}
\affiliation{
School of Electrical Engineering and Telecommunications, The University of New South Wales, Sydney, NSW 2052, Australia
}

\author{\\Arne Laucht}
\email[]{a.laucht@unsw.edu.au}
\affiliation{
School of Electrical Engineering and Telecommunications, The University of New South Wales, Sydney, NSW 2052, Australia
}

\author{Alexander Solntsev}
\affiliation{
School of Mathematical and Physical Sciences, Faculty of Science, University of Technology Sydney, Ultimo, New South Wales, 2007, Australia
}

\author{Milos Toth}
\affiliation{
School of Mathematical and Physical Sciences, Faculty of Science, University of Technology Sydney, Ultimo, New South Wales, 2007, Australia
}
\affiliation{
ARC Centre of Excellence for Transformative Meta-Optical Systems, Faculty of Science, University of Technology Sydney, Ultimo, New South Wales, 2007, Australia
}

\author{Robert Malaney}
\affiliation{
School of Electrical Engineering and Telecommunications, The University of New South Wales, Sydney, NSW 2052, Australia
}

\author{Igor Aharonovich}
\email[]{Igor.Aharonovich@uts.edu.au}
\affiliation{
School of Mathematical and Physical Sciences, Faculty of Science, University of Technology Sydney, Ultimo, New South Wales, 2007, Australia
}
\affiliation{
ARC Centre of Excellence for Transformative Meta-Optical Systems, Faculty of Science, University of Technology Sydney, Ultimo, New South Wales, 2007, Australia
}

\begin{abstract}
Quantum Key Distribution (QKD) is considered the most immediate application to be widely implemented amongst a variety of potential quantum technologies. QKD enables sharing secret keys between distant users, using  photons as information carriers. \al{An ongoing endeavour is to implement these protocols in practice in a robust, and compact manner so as to be efficiently deployable in a range of real-world scenarios.}
Single Photon Sources (SPS) in solid-state materials are prime candidates in this respect. Here, we demonstrate a room temperature, discrete-variable quantum key distribution system using a bright single photon source in hexagonal-boron nitride, operating in free-space. \al{Employing an easily interchangeable photon source system}, we have generated keys with one million bits length, and demonstrated a secret key of approximately 70,000 bits, at a quantum bit error rate of 6\%, with $\varepsilon$-security of $10^{-10}$. \al{ Our work demonstrates the first proof of concept finite-key BB84 QKD system realised with hBN defects.}
\end{abstract}

\maketitle

\section{\label{sec:introduction} Introduction}

Secure and hacking-proof communications is a vital requirement in today’s world. Traditional public key cryptography relies on lengthy and hard to decipher mathematical functions to encrypt and decrypt data. However, with the advancements of quantum computers, secured communication is increasingly vulnerable to hacking attempts. Quantum Key Distribution (QKD)~\cite{diamanti2016practical,xu2020secure,vajner2022quantum} the best-known application of quantum cryptography, offers an information-theoretic secure communication system, largely on account of the quantum non-cloning theorem~\cite{scarani2009security}. It enables two users to generate the exact same key without sharing any part of it publicly, providing a solution to secure key exchange. Since its inception in mid-1980s, and the first successful proof-of-concept test~\cite{Bennett_2014}, QKD has evolved to include various approaches such as entanglement-based QKD~\cite{yin2020entanglement, basso2021quantum, schimpf2021quantum, basso2022daylight}, measurement-device-independent QKD~\cite{lo2012measurement}, quantum teleportation-based QKD~\cite{bashar2009review}, and satellite-based QKD~\cite{liao2017satellite}. So far the majority of QKD systems rely on either nonlinear down-converted sources or attenuated lasers~\cite{ma2008quantum,kupko2020tools}. The advantage of these latter sources is their potentially high repetition rate. One alternative approach is using a deterministic, triggered single photon source (SPS) that emits a single photon per excitation cycle. 

Over the last few decades, significant effort has been put forward to develop such sources, with prime challenges being their purity (i.e., minimisation of multiphoton events) and extraction of light (i.e., collection efficiencies)~\cite{senellart2017high, aharonovich2016solid}. While semiconductor quantum dots are a great choice for a bright and pure source~\cite{basso2021quantum, heindel2012quantum, takemoto2015quantum,morrison2022single,bozzio2022enhancing}, their operation is limited to cryogenic temperatures. For wide deployment and practical implementation of QKD in real-world settings, compact, room temperature, sources are required~\cite{samaner2022free, leifgen2014evaluation, piparo2017measurement,murtaza2022efficient}.
Among the various solid-state materials, single-photon sources in hexagonal boron nitride (hBN) are considered a prime candidate for QKD owing to the material's favourable physical and optical properties~\cite{aharonovich2022quantum}. In particular, properties such as a high single-photon purity, and high-brightness operating in ambient conditions give a competitive advantage over other sources~\cite{zeng2022integrated}, \al{and a demonstration of the B92 QKD protocol has already been reported in Ref.~\onlinecite{samaner2022free} with a sifted key rate of 238 bit/s (similar to our raw key and bounded raw key rates below) and quantum bit error rates (QBERs) of 8.95\%}. 

After having demonstrated the in-principle usability of hBN SPSs for QKD in Ref.~\onlinecite{zeng2022integrated}, in this work, we include a full implementation of a free-space, discrete-level QKD system using an integrated SPS in hBN. We implement the BB84 protocol, demonstrating the sending, receiving and encryption/decryption process of an image from one device to another. We perform all security protocols, including privacy amplification, to demonstrate the most reliable QKD realised with SPSs to date.

\section{Setup Configuration}

\begin{figure*}[ht]
\includegraphics[width=0.8\textwidth]{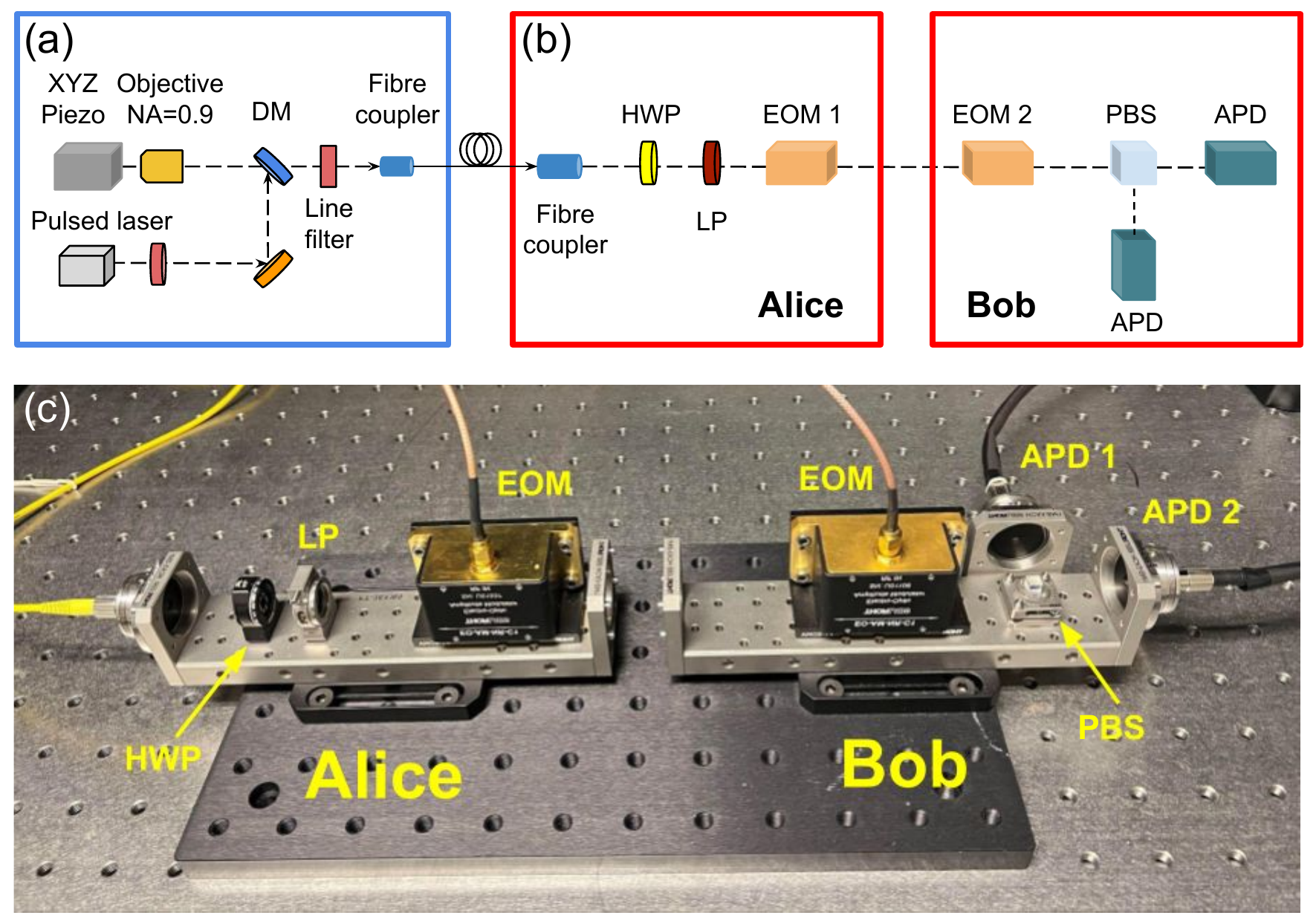}
\caption{(a) Schematic diagram of the single-photon source. DM, dichroic mirror. (b) Schematic diagram of the QKD setup. EOM, electro-optic modulator; LP, linear polariser; APD, avalanche photodiode; PBS, polarising beam splitter; HWP, half-wave plate. (c) The optical components of the transmitter (Alice) and the receiver (Bob).}
\label{figure_1}
\end{figure*}

The optical apparatus used to perform the measurements as a whole can be split into two main systems; the source to generate the single photons, and the QKD apparatus to perform the key distribution measurements. The hBN single photon sources are the primary source of photons for the QKD system. The sources are integrated with a solid immersion lens and packaged into a compact, portable device as described previously~\cite{zeng2022integrated} and shown schematically in \cref{figure_1}a. For the QKD experiments, the sources are excited using a 515 nm pulsed laser (PicoQuant PDL-800D), and the collection is coupled via single mode optical fibre to the QKD sender (Alice).

The QKD-side setup is shown in \cref{figure_1}b. The management and distribution of keys are performed by elements which modulate and/or measure the polarisation of the photons. The elements lie in a transmitter and receiver, termed Alice and Bob, respectively, that are separated by tens of centimetres in free space. The signal from the source is sent to the Alice, firstly. Alice consists of a half-wave plate (FBR-AH1, Thorlabs), linear polariser (FBRP, Thorlabs), and an electro-optic modulator (EOM) (EO-AM-NR-C1, Thorlabs). The half-wave plate and linear polariser are used to align and filter vertically polarised photons, respectively, minimising losses as they enter the EOM – a requirement for its function. The EOM is driven by a digital-to-analogue converter (DAC) (EVAL-AD5754REBZ, Analog Devices), multiplexer (MUX36D04EVM-PDK, Texas Instruments), and a high voltage amplifier (HVA 200, Thorlabs), supplying the four voltages (two voltages in Bob’s case) to the EOMs to induce the desired polarisation states.  Bob consists of an EOM, polarising beam splitter (PBS052, Thorlabs), and two avalanche photodiodes (APD, SPCMAQRH-12-FC, Excelitas).

\section{Experimental Procedure and Results}

\begin{figure*}[htb]
\includegraphics[width=0.8\textwidth]{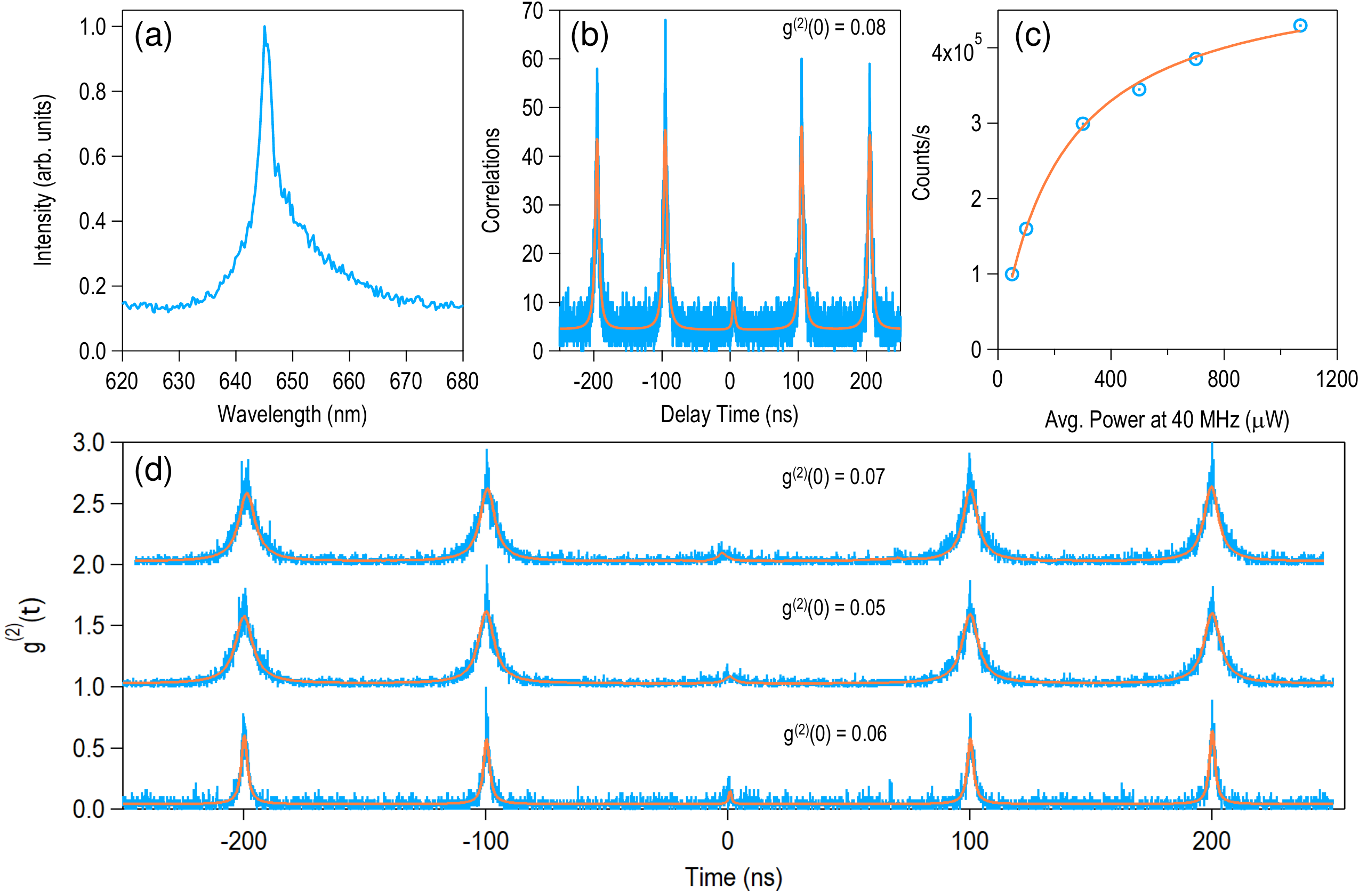}
\caption{Optical characteristics of the hBN SPS as measured through the SIL. (a) Spectrum of the SPS used for QKD, showing a sharp ZPL at 645 nm. (b) Pulsed second-order autocorrelation function of the QKD SPS at 10 MHz, showing a $g^{(2)}(0) = 0.08$. \al{The background in this measurement originates from the dark counts of our detectors.} (c)  Power-dependent saturation measurements showing count rate as a function of pulsed excitation power at 40 MHz repetition rate. (d) Pulsed second-order autocorrelation measurements of three characteristic SIL-integrated hBN SPSs, showing typical single-photon purity. All measurements were performed at 10 MHz repetition rate.}
\label{figure_2}
\end{figure*}

We start by choosing the most appropriate hBN SPSs – in particular, those with a high single-photon purity. \cref{figure_2}a shows the spectrum of the hBN SPS used in all QKD runs reported here.
This SPS has a characteristic sharp zero-photon line (ZPL) at 645 nm \al{(see \cref{figure_2}a), with an intrinsic linear polarization of 95.6\%.} The emission was then bandpass filtered (Semrock, $650\pm13$~nm) to select only photons from around the ZPL for the QKD experiment. The bandpass filtering was crucial in overcoming the wavelength dependence of the EOM and ensuring that the single-photon purity remains as low as possible, with \cref{figure_2}b showing the second-order autocorrelation function $g^{(2)}(0) = 0.08$. \al{From the $g^{(2)}(t)$ measurement, we can also extract the radiative emission time of $\tau=3.87$~ns.}

Additionally, the count rate was measured as a function of \al{average} pulsed excitation power at a pulse repetition rate of 40 MHz, as shown in \cref{figure_2}c. \al{By fitting the curve to $I = I_{\rm Sat}[P / (P + P_{\rm Sat})]$, we find the SPS to saturate at a power of $P_{\rm Sat} = 217$~$\mu$W, with a count rate of $I_{\rm Sat}=5.08\times10^5$ cts/s} as out-coupled from the single-mode fibre. From this, the calculation of the source-side total setup efficiency can be quantified with a mean photon number per pulse of $\mu = 0.012$. \al{This value was not corrected and includes contributions from single-photon emitter collection, optical losses, and detector efficiency - giving a realistic characterisation of the entire system. The number of photons be pulse} can be substantially improved by optimising the collection further, employing different designs~\cite{chen2011_99, sortino2021bright}. \al{Our experimental setup enables rapid characterisation and swapping of other sources, and we show examples of three other suitable sources with low $g^{(2)}(0)$ values in \cref{figure_2}d.} 

\begin{figure*}[htb]
\includegraphics[width=0.8\textwidth]{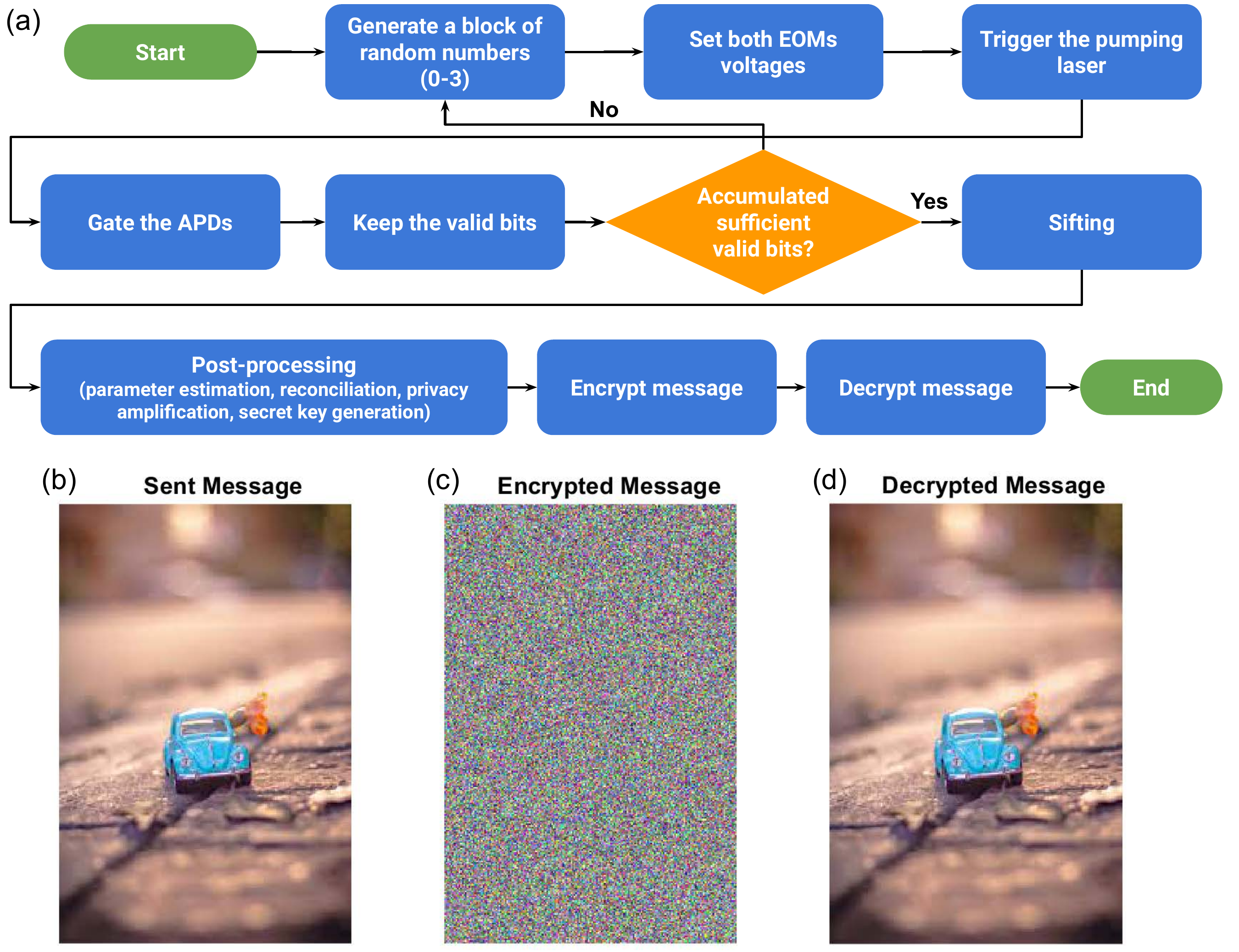}
\caption{QKD operation sequence. (a) Flowchart of the QKD process. (b) Original image, (c) image encrypted with Alice’s secure key, and (d) the decrypted image after decoding it using Bob’s secure key.}
\label{figure_3}
\end{figure*}

Having characterised and established the source of single photons, the testbed for the QKD system was then initiated. \cref{figure_3}a shows the operation sequence of the QKD process. The process starts with Alice generating a sequence of random integer numbers, each of which is 0, 1, 2 or 3, while Bob generates a sequence of random numbers between 0 and 1~\footnote{\al{The random number generation happens in software. In an actual deployment this software implementation will be replaced by an embedded quantum random number generator.}}. Alice’s numbers are used to encode the measurement bases and bit values onto the photons, while Bob’s numbers are used to select the measurement bases for the measurements. 

More specifically, Alice’s numbers map to one of four polarisation states imparted to the photons, horizontal (H,$\leftrightarrow$), vertical (V,$\updownarrow$), right-handed circular (R,$\circlearrowright$), or left-handed circular (L,$\circlearrowleft$). For each laser pulse cycle, at a rate of 500~kHz, Alice’s randomly generated number switches a multiplexer to select a specific one of four channels from the digital-to-analogue converter (DAC) to apply a predetermined voltage to the EOM and, as such, encodes the desired polarisation state onto the photon. 

Bob uses a corresponding setup to randomly select a basis, H/V or R/L, and measure the photon to obtain one of four outcomes. Bob’s EOM either leaves the incoming photon in its polarisation base or interchanges linear and circular polarisation. The PBS (polarising beam splitter) then separates linearly polarised photons with certainty to be detected at one of two detectors, whereas circularly polarised photons are unpredictably registered at either one of the two detectors. Bob’s APDs are gated to the laser pulse \al{with a gating time of 40~ns}, \al{and only the detection event of a single photon by one of the APDs per time bin is considered a valid bit and stored to the computer.
}


\begin{table*}[htp]
\begin{center}
\begin{tabular}{l|R{1.5cm}|R{1.5cm}|R{1.5cm}|R{1.5cm}|R{1.5cm}|R{1.5cm}}
\textbf{Experiment no.} & \textbf{1} & \textbf{2} & \textbf{3} & \textbf{4} & \textbf{5} & \textbf{6}\\
\hline
\hline
\al{Clock rate (Hz)} & 500,000 & 500,000 & 500,000 & 500,000 & 500,000 & 500,000\\
Photon detection rate at Bob (cts/s) & 1527 & 1486 & 1375 & 1722 & 1666 & 1001\\
\hline
Raw key length (bits) & 1,000,000 & 1,000,000 & 100,000 & 100,000 & 10,000 & 10,000\\
Raw key rate (bits/s) & 88 & 96 & 80 & 100 & 96 & 56\\
Bounded raw key rate (bits/s) & 396 & 432 & 360 & 450 & 432 & 252\\
QBER & 0.07 & 0.06 & 0.08 & 0.05 & 0.03 & 0.08\\
\end{tabular}
\end{center}
\caption{Selected experimental runs with their respective parameters.}
\label{table_1}
\end{table*}%

After revealing their bases used, Alice and Bob “sift” their events, discarding those which have mismatched bases. By defining H and R to correspond to a bit value of “0”, and V and L to correspond to a bit value of “1”, respectively, Alice and Bob retain a partially-correlated string of random bits. This string must be processed further to obtain a \textit{secret key}.

The required number of bits in the partially-correlated string must be large enough to provide a non-zero secret key rate at the pre-assigned $\varepsilon$-security (total failure probability)~\cite{scarani2008quantum,cai2009finite,chaiwongkhot2017finite,chaiwongkhot2020enhancing}, as discussed in the appendix. After accumulating the required number of bits, the post-processing procedure is initiated, a first component of which is parameter estimation. After this, a next phase is commenced, which consists of reconciliation and privacy amplification, after which the final secret key is generated.

In the QKD experiments we transmitted blocks of 1 million random bits at a clock rate of 500~kHz (limited by the voltage amplifier and the EOM), resulting in a 2~s transmission phase during which Bob received photons at a detection rate of $\approx1500$~cts/s. After this transmission phase, our system required 7~s for the data processing phase (dominated by the data transfer between the FPGA board and the PC), resulting in a total of 9~s per block of 1 million random bits. During the experiments no. 1 and no. 2 (in \cref{table_1}) we repeated the transmission of these blocks until 4 million valid bits were accumulated (see also flow chart in \cref{figure_3}a). 50\% of the accumulated bits are sacrificed during the sifting process, while another 50\% of the remaining bits are used for the quantum bit error rate (QBER) estimation. The partially-correlated keys that exist at this phase are referred to as the `\textit{raw keys}'.  Error correction is then \al{initiated} to turn the partially-correlated pair of keys into identical keys. Following this, the now correlated keys are input to the privacy amplification process of which, approximately $1/16^{\rm th}$  end up as the final secret keys (see appendix for details on error correction and privacy amplification).

We repeated the QKD measurement numerous times, accumulating raw keys of lengths up to 10$^6$ bits (after sifting and QBER estimation). Representative results for several independent runs of different lengths are shown in \cref{table_1}, with QBER values ranging between 3\% and 8\%. \al{We attribute the variation in QBER values to changes in the experimental conditions, such as drifts in the gain of the high voltage amplifiers (HVA200, Thorlabs) and background light in the room.}

Our experimental set up was primarily aimed at the development of the most reliable key in terms of $\varepsilon$-security, not the key rate. Hardware limitations imposed certain constraints on us, related to the amount of quantum information that could be transferred before classical protocols were implemented (e.g., TCP socket connections, memory-read functions), which led to time delays of 7~s per million bit-transfer attempts. Of course, such processing delays can never be set to zero in any implementation, but if we simply set all such delays to zero, we derive what we refer to as the \textit{bounded raw key} rate in \cref{table_1}. Beyond the time delays, the three most important factors influencing the raw key rate are (i) the input pump rate; (ii) capture rates from the photon source; and (iii) losses in our optical components. Achievable improvements in these three factors alone would readily lead us into the $\sim 10$~kHz raw key rate range.


For runs no. 1 and no. 2 (in \cref{table_1}) we also performed privacy amplification at $\varepsilon$-security levels of 10$^{-10}$ and below, and achieved secret key lengths of 33176~bits \& 68516~bits at secret key rates of 4~(bits/s) \& 6 (bits/s) or bounded secret key rates of 18 (bits/s) \& 26 (bits/s), respectively. For the other runs in \cref{table_1}, finite key effects  result in zero secret key rates at a $\varepsilon$-security level of 10$^{-10}$.

Finally, to demonstrate the QKD utility, we transfer an image in a secured way from Alice to Bob. The image of a toy car consisting of 48 kbits is shown in \cref{figure_3}b. We use the key from the QKD experiment no. 2 (see \cref{table_1}) as a one-time keypad to encrypt the image (\cref{figure_3}c). Alice then transmits the image to Bob classically, and Bob decrypts the image using the secure key at his end, as shown in \cref{figure_3}d. 

\al{A direct comparison with other reported single-photon QKD secure key rates is not straightforward given the difference in assumptions used (finite vs asymptotic, security equations, etc.) and technology used across the different reported works. Nonetheless, it may still be of value to attempt some form of key-rate comparison. We believe the raw key rate is best in this regard as most security assumptions have no impact on this rate.} 

\al{In the prototype reported here we found a raw key rate of 100 bits/s at 0.5 MHz clock rate. This is almost comparable to that reported in Ref.~\onlinecite{basso2021quantum} using a GaAs quantum dot (106 bits/s at 320 MHz clock rate), a factor $40\times$ less than Ref.~\onlinecite{leifgen2014evaluation} using the NV in diamond (3.99 kbits/s at 1 MHz clock rate), a factor $100\times$ less than Ref.~\onlinecite{rau2014free} using an electrically driven InAs quantum dot (5--17 kbits/s at 125 MHz clock rate), and a factor $270\times$ less than that reported in Ref.~\onlinecite{heindel2012quantum} using InP and InAs quantum dots (27--35 kbit/s at 182.6 MHz clock rate). Furthermore, recent experiments have performed QKD with frequency-converted quantum dot SPSs over longer distances, achieving 1--689 kbits/s depending on fibre length at 160 MHz clock rate in Ref.~\onlinecite{morrison2022single} and 5--6 kbits/s at 72.6 MHz clock rate in Ref.~\onlinecite{zahidy2023quantum}.}

\al{We caution again, that direct comparison of rates across different experiments, even when just raw key rates, is not straightforward. Many factors, including clock rates, processor speeds, sifting protocol used, and fractions sacrificed for error estimation are in play. Additional issues are raised when secure key rates are to be compared. Care must be taken.}

\section{Discussion and Outlook}

We report a functional, free-space, room-temperature QKD prototype with high-purity hBN SPSs. We implement a complete BB84 protocol, including privacy amplification and QBER corrections at a security level (total failure probability) of 10$^{-10}$. As discussed further in the appendix, although our rates are lower relative to others seen in the literature~\cite{leifgen2014evaluation, gao2022quantum, rau2014free}, straightforward comparison of the many published results should be done with caution as many different assumptions and security settings can be in place. In addition, our clock rate operates at 500~kHz, compared to other experiments that trigger at a few MHz rates, or faster. 

Furthermore, we have included all possible assumptions in our derivation of security in the finite key limit. \al{We have assumed the photons from our source contain no significant vacuum contributions, and therefore the issues raised in Refs.~\onlinecite{grunwald2019effective,chavez2020estimating} are neglected.} We believe our experimental secret key rates currently represent the most reliable in terms of security for the type of photon source used. Increasing the rate at a given security level, can be achieved via various improvements in the system – including an increased photon collection rate from the SPS, increased speed in the electronics, and increased computational power for the classical reconciliation. For instance, one key limitation originates from the high voltage amplifiers needed to drive the EOMs, which can be improved using alternative hardware. Using different field-programmable gate array (FPGA) hardware and data transfer protocols, can also further increase the rates.

All in all, our source has one of the best purities at room temperature, with an excellent brightness, and operation under ambient conditions that \al{promises} a straightforward integration with existing QKD networks. Our work paves the way for scalable implementation of QKD systems and holds great promise for using triggered room temperature SPSs based on hBN. 

\section*{Acknowledgements}
The authors thank Timm Kupko, Tobias Heindel, Oliver Benson, Esteban Gómez-López, Çağlar Samaner, Serkan Ateş, Maja Colautti, and Francois Ladouceur for useful discussions. We also thank Rich Mildren and Adam Bennett for the initial assistance of the source integration. This work is supported by the Australian Research Council (CE200100010), and the Office of Naval Research Global (N62909-22-1-2028). We thank the NSW Defence Innovation Network, the NSW State Government, and the Next Generation Technologies Fund for financial support of this project funded by the DIN Strategic Investment Initiative. \al{Lastly, the authors thank the two anonymous referees for their helpful comments that improved the quality of the manuscript.}

\newpage
\onecolumngrid
\appendix
\section{Reconcillation}

Reconciliation is a classical post-processing step that corrects the discrepancies between Alice and Bob's bit strings. Let us assume the estimated QBER, $Q$, is already achieved (via the prior sacrifice of $m$ bits) and that Alice and Bob start reconciliation with two bit strings (the raw keys), $K_{\rm A}$ and $K_{\rm B}$, each of length $n$ bits.  
The common treatment is to break $K_{\rm A}$ and $K_{\rm B}$ into shorter sub-blocks and then reconcile them in parallel~\cite{guo2021100}. Applying this treatment to our experiments,  the procedure of reconciliation is as follows:

\begin{itemize}
   \item \textbf{Step 1: Partition.} Alice breaks her $K_{\rm A}$ into multiple sub-blocks and the size of each sub-block is set to $n_{\rm block}$ ($n_{\rm block} \leq n$).  Bob does the same to his $K_{\rm B}$. Then, Alice and Bob will apply Step 2 and 3 to each of their sub-blocks. In our experiments we set $n_{\rm block}=10^4$, and note  that a maximum of 8 sub-blocks can be simultaneously reconciled due to the limited number of threads available at Alice.
    
   \item \textbf{Step 2: Syndrome Calculation.} Bob applies an LDPC matrix, $\mathbf{H}$, to his sub-block and obtains syndrome bits. Then, Bob sends the syndrome bits, $S_{\rm B}$ (of length $s_{\rm B}$), to Alice via classical communications. 
   We adopted an LDPC code designed to maximise its decoding threshold at a code rate, $R_{\rm c} = 0.5$. [Note, the decoding threshold of a given LDPC code is the maximal bit error rate so that a belief-propagation decoder is guaranteed to correct all the errors in an LDPC block~\cite{richardson2001design} - it can be maximised by using \emph{Density Evolution}~\cite{richardson2001design}.]
   The degree distribution polynomials of the adopted code can be found in Table~I of~\cite{elkouss2009efficient}. The matrix $\mathbf{H}$ (with $n_{\rm block}(1-R_{\rm c})$ rows and $n_{\rm block}$ columns) was constructed by the Progressive Edge Growth (PEG) algorithm~\cite{hu2005regular} based on the above mentioned polynomials.
     
    \item \textbf{Step 3: Decoding.} Alice uses Bob's syndrome bits, her sub-block, $Q$, and $\mathbf{H}$ as inputs to her LDPC decoder to correct all the discrepancies between Alice and Bob's sub-block. 
    The LDPC decoder used  is a serial-scheduled belief propagation decoder~\cite{sharon2007efficient} implemented in C++.

    \item \textbf{Step 4: Reorganising.} After all the sub-blocks are reconciled, Alice reorganises all her sub-blocks into a single bit string (the reconciled key), $\hat{K_{\rm A}}$, again of length  $n$ (similarly Bob to get $\hat{K_{\rm B}}$). 
   
\end{itemize}

Alice and Bob then proceed to privacy amplification to generate two identical and secure keys for cryptography purposes.  
Let us define the ratio, $r= s_{\rm final}/n$, where $s_{\rm final}$ is the final key length required to achieve a set security level. To determine $r$
we consider the security analysis in the finite-key length regime in Refs.~\onlinecite{scarani2008quantum,cai2009finite,chaiwongkhot2017finite,chaiwongkhot2020enhancing}. Let us further define $\varepsilon$-\emph{security} as the total failure probability, $\varepsilon$, of the protocol. Specifically, $\varepsilon = \Tilde{\varepsilon} + \varepsilon_{\rm PA} + \varepsilon_{\rm EC}+\varepsilon_{\rm PE}$, where $\Tilde{\varepsilon}$ is the smoothing parameter for the smooth min-entropy calculation, $\varepsilon_{\rm PA}$ is the failure probability of privacy amplification, and $\varepsilon_{\rm EC}$ is the failure probability of error correction. The probability $\varepsilon_{\rm PE}$ is somewhat more involved~\cite{scarani2008quantum,cai2009finite,finitetheory}. Consider the key, $K_{\rm B}^N$, held by Bob (similar to $K_{\rm B}$ but including the $m$ states that are to be sacrificed, i.e., $N=n+m$), and the key, $E^N$, held by the eavesdropper. Defining the combined quantum state held by Bob and the eavesdropper as
${\rho _{{K_{\rm B}^N}{E^N}}} = {\left( {{\sigma _{{K}{E}}}} \right)^{ \otimes N}}$, where $K$ and $E$ represent individual components of ${K_{\rm B}^N}$ and ${E^N}$, respectively, then $\sigma$ is contained in the set $\Gamma_\xi$ defined by 
$     \{ \sigma: \lvert\lvert \lambda_m - \lambda_\infty(\sigma) \rvert\rvert \leq \xi \}\,,$
except with  probability $\varepsilon_{\rm PE}$. Here,
 $\lambda_m$ are the statistics derived from measurements on $m$ samples of $\sigma$; 
 $\lambda_\infty(\sigma)$ is the probability distribution defined by the measurements on $\sigma$; and  $\xi=\frac{1}{2}\sqrt{\frac{2\log_2(\frac{1}{\varepsilon_{\rm PE}})+d\log_2(m+1)}{m}}$,  where $d=2$ is set due to the positive operator valued measure with 2 outcomes.
 These  expressions are used to form an upper limit, $Q^u$, to $Q$ that is used in determining the final key length. In many QKD variants this takes the simple form
$Q^u=Q+\xi$. In more simpler terms, we can set an upper limit to what we think the true QBER, $\hat{Q}$, is and then determine the probability that this upper bound does not encompass $\hat{Q}$. For our protocol implementation, we  use a similar analysis to \cite{cai2009finite} but with equal probability for each basis usage. Parameter estimation from individual bases was investigated, but in our experiments the QBER from each basis was found to be the same.

 All of the above discussion  leads us to a  relation for $r$  given by \footnote{\al{A typographical error in~\cite{chaiwongkhot2020enhancing} (and carried over into~\cite{zeng2022integrated}) is corrected for here.}} 
    \begin{equation}
    \label{eq:keyLength}
    \al{
        r = \frac{s_{\rm final}}{n} =  A\left(1-h\left(\frac{Q^u}{A} \right)\right) - (1-R_{\rm c}) - \Delta(n)\,,
        }
    \end{equation}
where $A = \frac{p_{\rm det} - P_{\rm m}}{p_{\rm det}}$, is the correction term due to the multi-photon emission at the single-photon source. In this latter relation, $p_{\rm det}$ is the probability of detecting at least one photon, $P_{\rm m}$ is the probability of multi-photon emission at the source 
\al{We obtain a determination of $A$ from a direct measurement of the ratio of $P_m/p_{det} = 0.015$, leading to
$A = 0.985$.}

In Eq.~(\ref{eq:keyLength}), $h(x) = -x\log_2 x - (1-x)\log_2(1-x)$ is the binary entropy function, and $\Delta(n)$ is an additional penalty term  given by
\begin{equation}
\label{eq:delta}
    \Delta(n) = \frac{7n\sqrt{\frac{1}{n}\log_2 \frac{2}{\Tilde{\varepsilon}}} + 2\log_2\frac{1}{\varepsilon_{\rm PA}} + \log_2 \frac{2}{\varepsilon_{\rm EC}}}{n} \,.
\end{equation}
We use the term $(1-R_{\rm c})$ to compute the fraction of information leakage during reconciliation instead of the commonly used term $f_{\rm E}h({Q})$, where $f_{\rm E}$ is the reconciliation efficiency. Noting  that only the syndrome bits are disclosed during reconciliation, this efficiency can be written   ~\cite{elkouss2011information} 
 $   f_{\rm E} = \frac{s_{\rm B}}{nh({Q})}=\frac{1-R_{\rm c}}{h({Q})}\,,$
since $\frac{n- s_{\rm B}}{n} = R_{\rm c}$ holds for a given LDPC matrix.

\begin{figure}[tb]
\centering
\includegraphics[width=0.7\linewidth]{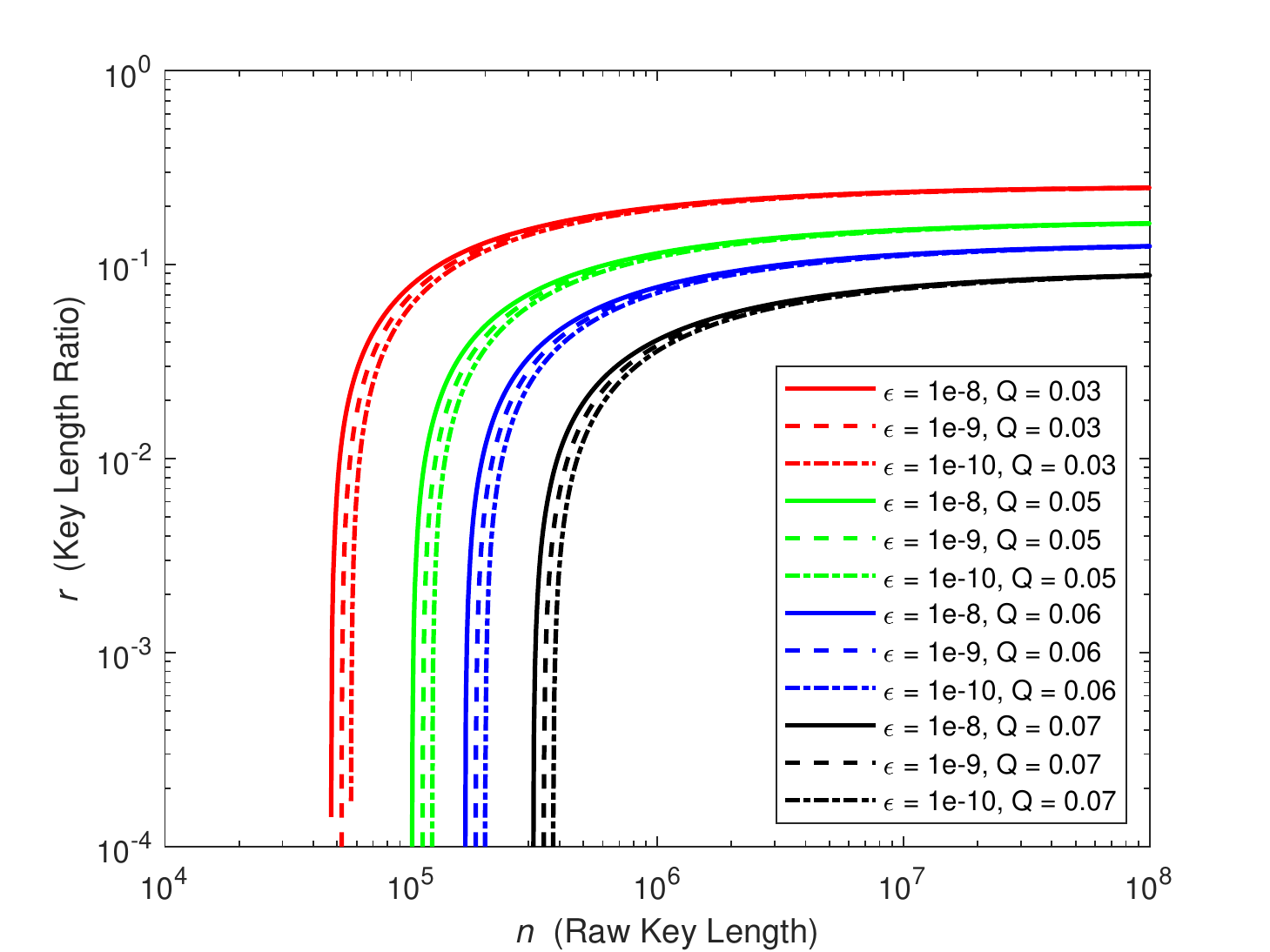}
\caption{The ratio of key lengths, $r$, vs. the raw key length, $n$, for different security levels and different $Q$ values. Here, the number of bits used for parameter estimation, $m$, is set at $m=n$.}
\label{plot3Q}
\end{figure}

In the main text we investigated the achievable $r$ with respect to $\varepsilon=10^{-10}$ with the results reflected in the main text. In all our reported secret key rates, we have set $m = n =10^6$ to ensure a non-zero secret key rate at our required security level.  
 
We note that our achievable $r$ is a lower bound and can be further improved by optimising some of the parameters (e.g., $\Tilde{\varepsilon}$) under defined constraints on the probabilities. Such optimisation was not carried out in our results.
 We have fixed $\Tilde{\varepsilon} = \varepsilon_{\rm PA} = \varepsilon_{\rm EC} = \varepsilon_{\rm PE} = \frac{\varepsilon}{4}$.
Typical examples of $r$ as a function of $n$ are shown in Fig.~\ref{plot3Q}.

\al{Although an emphasis was put on the inclusion of all known effects impacting the derived security level, we do remind the reader that all BB84 implementations come with a host of assumptions  on the anticipated behavior of devices, including the lack of side-channel attacks. That is, BB84 security is device-dependent. The ultimate QKD security, achievable via device-independent QKD, requires entangled photon sources and measurements of Bell violations.}

We close by outlining how our privacy amplification is actually implemented.
\begin{itemize}
    \item \textbf{Step 1: Calculating $r$.} Alice uses Eqs.~\ref{eq:keyLength} and ~\ref{eq:delta} to obtain $r$ based on $Q$, $\varepsilon$, $\Tilde{\varepsilon}$, $\varepsilon_{\rm PA}$, $\varepsilon_{\rm EC}$, $\varepsilon_{\rm PE}$, $n$ and $m$. 
    
    \item \textbf{Step 2: Creating the hash function.} Following the procedure described in Section~II.E of~\cite{bourgoin2015experimental}, Alice creates a Toeplitz matrix, $\mathbf{T}$, with $n$ columns and $\lfloor r n\rfloor$ rows, where $\lfloor \cdot \rfloor$ is the floor operation. Alice sends $\mathbf{T}$ to Bob.
    
    \item \textbf{Step 3: Secure hashing.} Alice and Bob apply $\mathbf{T}$ to $\hat{K_{\rm A}}$ and $K_{\rm B}$, respectively, and obtain two identical and secure key strings for cryptography purposes. [At some points, an \emph{a priori} secret key will be consumed by Alice and Bob for authentication before the use of any key to encrypt/decrypt classical messages - we assume that such authentication is completed successfully.] A final check (hash) is taken on some small part  of the keys to check the keys are identical - if they are not the protocol is aborted.
\end{itemize}

\newpage
\twocolumngrid

\bibliography{bibliography}
\end{document}